# THE FCC-ee INTERACTION REGION MAGNET DESIGN

M. Koratzinos, A. Blondel, University of Geneva, Geneva, Switzerland; A. Bogomyagkov,
S. Sinyatkin, BINP SB RAS, Novosibirsk, Russia
M. Benedikt, B. Holzer, J. van Nugteren, F. Zimmermann, CERN, Geneva, Switzerland
K. Oide, KEK, Tsukuba, Japan

*Abstract*

The design of the region close to the interaction point of the FCC-ee [1] [2] experiments is especially challenging. The beams collide at an angle (±15 mrad) in the high-field region of the detector solenoid. Moreover, the very low vertical $\beta^*$ of the machine necessitates that the final focusing quadrupoles have a distance from the IP ($L^*$) of around 2 m and therefore are inside the main detector solenoid. The beams should be screened from the effect of the detector magnetic field, and the emittance blow-up due to vertical dispersion in the interaction region should be minimized, while leaving enough space for detector components. Crosstalk between the two final focus quadrupoles, only about 6 cm apart at the tip, should also be minimized.

## INTRODUCTION

FCC-ee incorporates a "crab waist" scheme to maximize luminosity at all energies [3]. This necessitates a crossing angle between the electron and positron beams which is currently assumed to be ±15 mrad in the horizontal plane. No magnetic elements can be present in the region approximately ±1 m from the interaction point (IP) to leave space for the particle tracking detectors and possibly the luminosity counter. Furthermore, the area outside the forward and backward cones of 100 mrad defined from the IP and along the longitudinal axis of the experiment is reserved for detector elements, leaving only the two narrow cones for machine elements. Therefore, beam electrons experience the full strength of the detector magnetic field in the vicinity of the IP. The resulting vertical kick needs to be reversed and this is performed in the immediate vicinity. This vertical bump, however, leads to vertical dispersion and an inevitable increase the vertical emittance of the storage ring. Since FCC-ee is a very low emittance machine (with an emittance budget of the order of 1 pm), the emittance blow-up in the vicinity of the IP needs to be minimized.

## THE MAGNETIC ELEMENTS AROUND THE IP

We now have a preliminary conceptual design of the magnetic systems close to the IP which fits our requirements. It comprises the following elements:

The *detector solenoid* is assumed to have a magnitude of 2 T and extend to ±6 m from the IP.

The *screening solenoid* is a thin solenoid producing a field equal and opposite to the detector solenoid and screens the final focus quadrupoles from the detector solenoid field. It starts at around 2 m from the IP and extends all the way to the endcap region of the detector.

The *compensating solenoid* sits in front of the screening solenoid, has a field higher than that of the detector solenoid, so that the magnetic field integral seen by the beam is zero. In our design the length of this solenoid is around 0.7m, and its strength is approximately 5 T.

The *final focus quadrupoles* in our current design sit at a distance of 2.2 m from the IP and are 3.2 m long. The focusing strength in the current design is 92 T/m at 175 GeV [4]. The distance between the centres of the two quadrupoles is 6.6 cm at the tip closest to the IP and 16.2 cm at the far end.

The different elements of the design can be seen in Figure 1, as seen from above the detector. Please note the elongated view along the x-axis. Figure 2 shows the magnitude of the magnetic field. All analysis described here was done using the *Field* suite of programs [5].

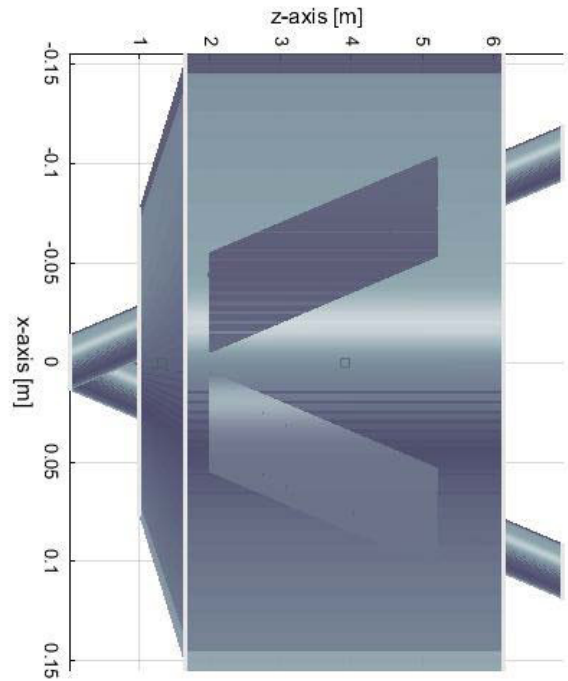

Figure 1: The conceptual design of the magnetic elements close to the IP, looking on the x-z plane. The IP is at (0,0). Please note the elongated scale in x. The opening angle of the (schematic) beam pipes is 30 mrad. The final focus quadrupoles surround the two beam pipes whereas the rest of the elements are aligned to the longitudinal axis of the experiment. The compensating solenoid is tapered and is in front of the screening solenoid. The detector solenoid is outside this picture.

## EMITTANCE BLOW-UP

The vertical emittance increase close to the IP, $\Delta\epsilon_{y,IP}$, is given by

$$\Delta\epsilon_{y,IP} = 3.83 \times 10^{-13} \frac{\gamma^2}{J_y} \frac{I_{5,IP}}{I_2} \quad (1)$$

Where $\gamma$ is the relativistic $\gamma$ of the beam, $I_2$ is the second synchrotron radiation integral which can be approximated by

$$I_2 \cong \frac{2\pi}{|\rho_{bend}|} \quad (2)$$

(equal to about $6 \times 10^{-4}$ for FCC-ee with bending radius in the arcs $\rho_{bend} = 11$ km. $J_y = 1$. The fifth synchrotron radiation integral is

$$I_{5,IP} = \int_{-d}^{d} \frac{\mathcal{H}_y(s)}{|\rho|^3} ds \quad (3)$$

where $\rho$ is the bending radius due to the magnetic field along the path of the electrons in the area of interest, $-d$ to $d$, in our case -3 to 3 m.

$$\mathcal{H}_y(s) = \beta(s){D'_y}^2 + 2\alpha(s)D_y D'_y + \gamma(s){D_y}^2 \quad (4)$$

where $D_y$ is the vertical dispersion and

$$\alpha(s) = -\frac{1}{2}\beta'(s); \gamma(s) = \frac{1+\alpha(s)^2}{\beta(s)} \quad (5)$$

Where $\beta(s)$ is the vertical beta optics function. Emittance blow up is worse at low energies due to the $\frac{\gamma^2}{|\rho|^3}$ dependence (the magnetic field of the detector is expected not to change at different energies).

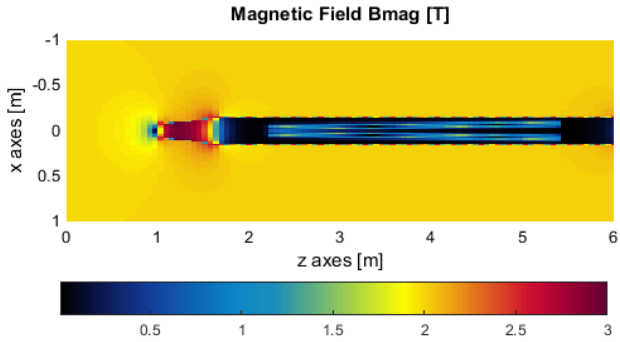

Figure 2: The magnitude of the magnetic field in the region x=(-1 m, 1 m) and z=(0, 6 m) in the vicinity of the compensating solenoid (red, -3 T), screening solenoid (black, 0 T), final focus quadrupoles (in blue), all in the +2 T solenoidal field of the experiment (yellow).

The minimization of the above formula for the emittance blow-up has been done empirically due to the number of parameters. The boundary conditions used were: detector solenoid field of 2 T, magnetic elements should be inside the 100 mrad forward and backward cones, location of closest element to the IP 1 m. The latter is certainly a tight requirement since the luminosity counter should sit in front of the first magnetic element. The size and position of the different components for the optimal case where the emittance blow-up was the smallest was as follows:

Compensating solenoid: inner edge at 1.0 m, length 0.65 m, diameter 16 to 22 cm (tapered). Current (1000 windings) 2615 A, giving a maximum field along the axis of -4.95 T

Screening solenoid: inner edge at 1.65 m, 2.5 m length, diameter 30 cm and current (10,000 windings) 717 A, giving a maximum field along the axis of -2 T.

The resulting emittance blow-up was computed using the SAD suite of programs [6] which gave a result of a total of 0.11 pm of vertical emittance blow up for two identical IPs. Optical functions can be seen in Figure 4.

## FINAL FOCUS QUADRUPOLES

The requirements for the final focus quadrupoles (here we refer to the last elements focusing the beam in y for a resulting $\beta_y^*$ of 1 mm) are as follows: the beam-stay-clear (b-s-c) area in the vicinity of the quadrupoles has been computed to be ±12 mm. This allows for a very compact beam pipe. RF beam heating considerations might necessitate a slightly larger beam pipe that the b-s-c suggests, as large as 40 mm diameter. The L* of these quadrupoles in the current design is 2.2 m and their strength is 92 T/m. Increasing the quadrupole strength would allow for shorter quadrupoles and therefore the L* requirement can be increased if needed. With a L* of 2.2 m, the distance between the centre of the quadrupoles is 6.6 cm, calling for a compact design and one that takes into account the cross talk between the two quadrupoles.

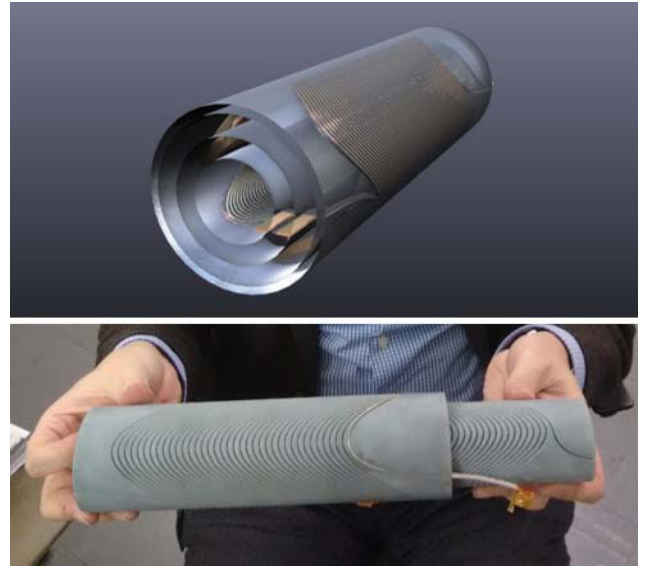

Figure 3: Prototype CCT final focus quadrupole. CAD drawing (top) and 3D printed item (bot).

One technology that looks promising for such an application is CCT technology.

Canted-cosine-theta (CCT) magnets have been around since the seventies [7], however only recently have they become popular with magnet designers [8] [9]. The CCT design offers some advantages over traditional magnet design for certain applications. The main advantages of CCT magnets are:

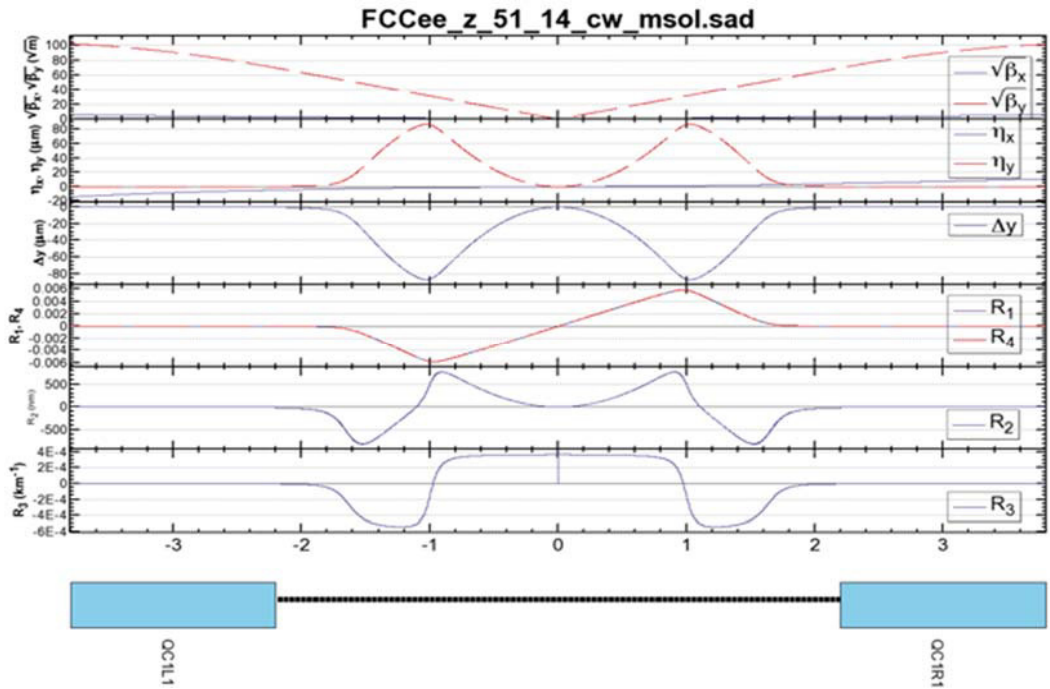

Figure 4: Beam optical functions between the centers of the final quadrupoles of the proposed solution. Columns are, from top to bottom, $\sqrt{\beta_x}$ and $\sqrt{\beta_y}$, x and y dispersion, y orbit, $R_1$ and $R_4$, $R_2$, $R_3$, where $R_{1,2,3,4}$ are the x-y coupling parameters. The vertical orbit, dispersion, and the coupling parameters are confined within the compensation solenoid.

- Accelerator-grade field quality
- Fast prototyping: short turnaround times using 3D printing techniques
- Easy to manufacture with state-of-the-art manufacturing techniques (CNC machines or 3D printing), simple windability and assembly; not labour intensive
- No need for coil pre-stress during assembly; also, reduced coil stresses should improve magnet training
- Total freedom to design any multipole arrangement, therefore capable of producing compact double aperture magnets with the required field quality
- Fewer components and considerably lighter than traditional designs – this might translate to reduced costs (although this currently has not been fully demonstrated)

All of the requirements of the FCC-ee final focus system can be satisfied using a CCT design for the final focus quadrupoles. The design is compact, has a theoretical field quality which is adequate for our stringent requirements, crosstalk can be designed out, and has the added bonus of fast prototyping which ensures fast progress, much greater than what is customary for magnet design. However, the design needs to prove that it can deliver a series of milestones, including adequate field quality and crosstalk correction capability.

FCC-ee is pursuing the CCT option for the final focus quadrupoles, while also keeping the more conventional modified Panofsky lens with twin-aperture and integrated iron yoke style design as an alternative. This design is undertaken at BINP.

## CCT PROTOTYPE

A complete conceptual design of an FCC-ee final focus prototype has been done (Figure 3). The prototype has the final dimensions regarding the bore size, but it is much shorter (20 cm) than the final magnet. The design has been 3D-printed in and awaits winding with an existing NbTi cable and will be ready to be measured for field quality and tested for cryogenic performance.

## CONCLUSIONS

We have demonstrated that the very stringent requirements for the magnetic systems around the IP of an FCC-ee detector can be met with a system comprising final focus quadrupoles, screening solenoid and compensating solenoid. The emittance blow-up due to two interaction regions is computed to be 0.11 pm, which is about 10% of our vertical emittance budget. Further improvements and a move to prototyping and technical design will follow.

## ACKNOWLEDGEMENTS

We would like to thank V. Telnov who was the first to bring to our attention the problem and its solution.